\documentstyle[a4]{article}

\begin{document}

\title{Kosterlitz-Thouless phase and continuous melting transition in layered 
superconductors immersed in a parallel magnetic field} 

\author{Xiao Hu and Masashi Tachiki\\
Computational Materials Science Center \\ 
National Institute for Materials Science,
Tsukuba 305-0047, Japan}


\date{(revised on April 24, 2003)}

\maketitle

\begin{abstract}
\(B-T\) phase diagram with a multicritical point for interlayer Josephson vortices is 
mapped out based on Monte Carlo simulations.  For high magnetic fields we find a novel
Kosterlitz-Thouless (KT) intermediate phase characterized by in-plane 
two-dimensional (2D), quasi long-range orders (QLROs) of vortex 
alignment and superconductivity.  Both the transition to high-temerature normal state
and the evolution to low-temperature phase of 3D LRO are continuous. Decoupling of the
3D system into the 2D state is triggered by hops of segments of Josephson flux lines
across superconducting layers activated by thermal fluctuations.
For low magnetic fields, a single first-order melting transition is observed.  

\vskip1cm


\end{abstract}

\newpage

The discovery of high-\(T_c\) superconductivity in cuprates has triggered extensive researches
into the finite-magnetic-field superconductivity transition in type II superconductors 
\cite{Review}. It is now well established that the transition is first order, accompanied by
the freezing of the flux-line liquid into lattice.  This notion is 
particularly important because the transition was considered second order for a long 
time since Abrikosov \cite{Abrikosov}.  

The high-\(T_c\) superconductors share a layered structure in which the superconductivity 
is widely believed to occur mainly in the CuO\(_2\) layers intervened by layers of charge 
reservoir.  This profound layer structure causes 
significant differences in both equilibrium and transport properties under magnetic fields
in different directions.  The first-order normal to 
superconductivity transition is mainly observed under magnetic fields perpendicular
to the layers, at which the two-dimensional (2D) translation symmetry enjoyed by pancake
vortices is broken. In a sharp contrast, a parallel magnetic field penetrates the system 
through the reservoir layers, in terms of Josephson vortices.  The relevant \(c\)-axis
translation symmetry is broken {\it a priori}, which may raise new phase and new melting
process. In fact, a peculiar transport phenomenon has been found in 
Bi\(_2\)Sr\(_2\)CaCu\(_2\)O\(_{8+y}\) by Iye et al.
\cite{Iye} that the resistivity does not depend on the angle between the 
magnetic field and current when they are both parallel to the CuO\(_2\) layer
\cite{Iye} and by Ando et al. that the system shows power-law, non-Ohmic
dissipations \cite{Ando}.  Chakravarty et al. pointed out that such a dissipation
cannot occur in a lattice phase \cite{Chakravarty}.  Blatter et al. \cite{Blatter} 
proposed a novel Kosterlitz-Thouless (KT) \cite{Berezinsky,KT} scenario 
at high magnetic fields, characterizing the behavior by a smectic state with vanishing
interlayer shear modulus (see \cite{Horovitz} for 
a possible KT phase at intermediate magnetic fields).  
Motivated by the experiment by Kwok et al. suggestive of continuous melting 
transition \cite{Kwok}, Balents and Nelson proposed a smectic phase in between the lattice and 
liquid states and thus a two-step, continuous melting scenario at intermediate magnetic fields
\cite{Balents} (see \cite{Radzihovsky} for a proposal of supersolid). Decoupling 
between superconducting layers was actually proposed even before the discovery of high-\(T_c\)
superconductors \cite{Efetov,Ivlev}.  Starting from a 2D elastic Hamiltonian and resorting to 
renormalization group (RG), however, it was argued by Mikheev and Kolomeisky 
\cite{Mikheev} that only a 3D long-range crystalline order is possible besides liquid state 
(see also \cite{Korshunov}).  The discrepancy among different approaches has not been 
resolved to provide a unified picture, though they provide new physical insights, 

By means of computer simulations we find in the present work that there is a 
multicritical point in the \(B-T\) phase diagram of interlayer Josephson vortices: 
Below the critical field, a single first-order transition upon temperature sweeping is
observed; above it, there exists an intermediate KT phase 
characterized by in-plane 2D, quasi long-range orders (QLROs) of vortex 
alignment and superconductivity in between the normal phase and 3D lattice phase,
accompanied by two continuous melting transitions. 

The Hamiltonian is for the phases of the superconductivity order parameter on 
the simple cubic lattice \cite{Hu1,Hu32}

\begin{equation}
  {\cal H}=-J\sum_{<i,j>\parallel x,y {\rm axis}} 
                     \cos(\varphi_i -\varphi_j) 
           -\frac{J}{\gamma^2}\sum_{<i,j>\parallel c {\rm axis}} 
                     \cos(\varphi_i-\varphi_j
                           -\frac{2\pi}{\phi_0}\int^{j}_{i}A_cdr_c),
\label{eqn:hamiltonian}
\end{equation}

\noindent where \(J=\phi^2_0 d/16\pi^3\lambda^2_{ab}\), the \(y\) direction along the 
external magnetic field and \(\hat{x}\perp \hat{c}\perp \hat{y}\), \({\bf A}=(0, 0, -xB)\), and 
\(\gamma=\lambda_c/\lambda_{ab}\).  The system is of size 
\(L_x\times L_y\times L_c=384d\times 200d\times 20d\) under periodic boundary 
conditions, with \(d\) the separation between CuO\(_2\) layers.  Vortices are identified by
counting gauge invariant phase differences \cite{Hu1}.
It was found \cite{Hu32} that the \(\delta\)-function peak in the specific
heat associated with the first-order melting of Josephson vortex lattice is suppressed
when the magnetic field and/or anisotropy parameter \(\gamma\) exceed the critical values
given by \(f\gamma=1/2\sqrt{3}\) (\(f\equiv Bd^2/\phi_0\)),
suggestive of a multicritical point and continuous melting(s) beyond it.   In the present
study, we fix \(f=1/32\) and take two typical anisotropy constants: 
\(\gamma=8\), realized in YBa\(_2\)Cu\(_3\)O\(_{7-\delta}\) and slightly
below the critical value for \(f=1/32\), and \(\gamma=20\), quite above
the critical value so that analysis on the low-temperature phase becomes easier.  

The first observation is made for the helicity modulus. As shown in Fig. 1, in-plane helicity 
moduli set up at \(T\simeq 0.96J/k_B\).   While the difference between 
the transition points of the two anisotropy parameters is very small to be seen clearly 
by the present system size, we differentiate them as \(T_m\) and \(T_{\rm KT}\) for 
\(\gamma=8\) and \(\gamma=20\), respectively, for the reasons elucidated later. The finite 
helicity modulus \(\Upsilon_y\) along the magnetic field 
indicates the phase coherence at low temperatures, corresponding to the breaking of
U(1) gauge symmetry and therefore appearance of superconductivity.  A finite value of 
\(\Upsilon_x\) is a reflection of the intrinsic layer pinning (that along the \(c\) axis
remains vanishing down to zero temperature).  The collapse of the helicity moduli in  
\(x\) and \(y\) directions for \(\gamma=20\) is an intrinsic property of the state the system
presumes beyond the multicritical point.  It is considered to be responsible to the experimentally
observed orientation independence of I-V characteristics, noting that the conductance can
be evaluated by helicity moduli.    

The second observation is on the density correlations of Josephson vortices. 
Structure factors \(S[k_x,y=0,k_c]\) at the transition points are displayed in Fig. 2.  
As seen in Fig. 2(a), six Bragg peaks appear for \(\gamma=8\) (see also Fig. 3), indicating 
that a 3D LR crystalline order is established upon the first-order freezing transition
(\(T_m\) refers to the melting point).  In a sharp contrast, the Bragg spots at 
\([\pm 2f\pi/d,\pm \pi/d]\) for \(\gamma=20\) shown in Fig. 2(b) are diffusive and stripe like.
As in Fig. 3, the \(k_c\) profile of the Bragg spots for \(\gamma=20\) is fitted well by
a Lorentzian function \(1/[\xi^{-2}_c+(k_c-\pi/d)^2]\) with the correlation length
\(\xi_c\simeq 0.47d\) in the \(c\) direction.  Josephson vortices are therefore decoupled 
into nearly independent layers at the transition point.  The correlation length grows to 
\(\xi_c\simeq 1.5d\) at \(T=0.7J/k_B\). The SR crystalline order in the \(c\) axis makes the
interlayer shear modulus vanishing, which was first discussed by Blatter et al. \cite{Blatter}. 

The \(k_x\) profiles of Bragg spots at \([k_x,k_c]=[\pm 2f\pi/d,\pm \pi/d]\) 
for \(\gamma=20\) are plotted in Fig. 4 at several typical temperatures.  Singularity
appears clearly in the \(k_x\) profile below \(T_{\rm KT}=0.96J/k_B\), consistent with the 
onset of the helicity moduli in Fig. 1.  By fitting the profiles of the Bragg spots to
the function \(I\simeq a|1-k_xd/2f\pi|^{\eta-2}+b|1-k_xd/2f\pi|+c\), denoted in Fig. 4 
by the solid curves, we estimate the exponent as \(\eta\simeq 2.40\pm 0.03, 2.21\pm 0.01\) and 
\(1.87\pm 0.01, 1.41\pm 0.02\) for \(T=0.96, 0.95\) and \(0.8, 0.7J/k_B\), with 
the error bars from the least-squares fittings; for \(T=0.92J/k_B\) a logarithmic function 
fits best with the data, corresponding to \(\eta=2\);  data for 
\(T=0.7J/k_B\) are not included for the sake of clarity. 
The power exponent \(\eta\) depends on temperature, 
implying that elastic constants are not simply entropy dominated as in polymer systems. 
The height of Bragg spots increases very slowly 
with decreasing temperature (see Fig. 1 of Ref.\cite{Hu32} for comparison).  
The Bragg spots become very sharp to be fitted by a power-law function for 
\(T\le T_{\times}\simeq 0.65J/k_B\);  there is no difference between the shape of structure
factors for \(\gamma=20\) at these low temperatures and that for \(\gamma=8\) in Fig. 2(a).
These observations indicate a 3D LR crystalline order realized even in highly anisotropic systems 
provided temperature is low enough.  

The singularities in the structure factors for \(\gamma=20\) at \(T_{\times}<T\le T_{\rm KT}\)
indicate unambiguously power-law density correlations, characterized by the exponent 
\(\eta\), along the \(x\) direction in real space.  
Combining with the similar observations along the 
direction of the magnetic field, with smaller exponents, a 2D QLR crystalline order is
concluded at these intermediate temperatures. Since all 2D QLROs associated with continuous 
degrees of freedom known to date are governed by the KT
fixed point, we identify the intermediate temperature region as a novel KT phase. 
The melting of this 2D Josephson vortex lattice at \(T_{\rm KT}\) is therefore considered as
a KT transition, at which the tilt modulus diminishes to zero.
Since the crystalline order is SR in \(c\) direction in the KT phase, 
so do the gauge invariant phase correlations.  The finite helicity moduli for 
\(\gamma=20\) below \(T_{\rm KT}\) correspond to QLROs of phase variables.

The third observation is on the trajectories of the Josephson flux lines. This is important 
since a decoupling between ordered layers is argued based on RG analysis to be impossible, 
providing Josephson flux lines are confined completely by the CuO\(_2\) layers \cite{Mikheev}.  
As depicted in Fig. 5, hops of Josephson flux-lines segments across CuO\(_2\) layers,
by creating pancake vortices,
into neighboring reservoir layers are observed for \(T> 0.6J/k_B\), and the
percentage of hopping Josephson flux lines increases quite sharply with 
temperature.   These observations move away the hurdle to the decoupling and the 2D phase,
and thus reconcile our simulation results with the RG analysis. 
Hops of Josephson flux-line segments disturb intralayer correlations,
weaken the interlayer correlations, as clearly seen in Fig. 4, and result in the suppression of 
3D crystalline order into decoupled 2D quasi lattices for weak bare Josephson couplings.  
For \(T\le T_{\times}\), hops of Josephson flux lines are suppressed, where
the argument by Mikheev and Kolomeisky should apply.  As a matter of fact, this temperature 
regime coincides roughly with that in which the 3D LR crystalline order becomes stable for
\(\gamma=20\).  In Fig. 5, we
also display the temperature dependence of populations of thermally excited vortex loops, 
and of collisions between Josephson flux lines residing in the same block layers.  These
results indicate clearly the importance of thermal fluctuations in the present
system, and the former explains why the exponent \(\eta\) depends on temperature.

The evolution from the 2D QLRO to 3D LRO at \(T_{\times}\) is also believed to be a 
thermodynamic phase transition.  Resorting to the effective Landau theory formulated 
first by Balents and Nelson \cite{Balents}, which works better for the present QLRO in 
the KT phase, it is argued that this phase transition is probably second order and in the 
3D XY universality class.  This observation can be taken consistent with the absence of
noticeable anomaly in the specific heat at \(T_{\times}\) \cite{Hu32,Olsson},
since the critical exponent \(\alpha\) is slightly negative 
in the universality class and thus the associated cusp could be very small.  The
interlayer shear modulus is suppressed continuously as \(C_{66}\sim (d/f\xi_c)^2\)
with an exponent \(2\nu\simeq 4/3\) as \(T_{\times}\) is approached from below. 

Possible size effects on our simulation results are addressed as follows.
The successful observation on the single, first-order melting transition for \(\gamma=8\)
indicates that the system size is sufficient for deriving right physics below the 
multicritical point.  Since increasing the anisotropy parameter only reduces the coupling 
in the \(c\) axis, the system size unlikely becomes insufficient above the multicritical 
point.  It is easy to see that the 2D QLRO phase \([T_{\times},T_{\rm KT}]\) does
not shrink to zero in the thermodynamic limit, since 
\(T_{\rm KT}>T_{\rm KT}^{\rm bare}>T_{\times}\), with 
\(T_{\rm KT}^{\rm bare}\simeq 0.89J/k_B\), is observed under periodic boundary conditions. 

Based on the analyses presented so long, we map out in Fig. 6 the \(B-T\) phase diagram 
for interlayer Josephson vortices, noting that the same physics should occur 
when the magnetic field is tuned while the anisotropy parameter is fixed. The first-order 
melting line for low magnetic fields branches into two phase boundaries at the multicritical 
point \(B_{mc}=\phi_0/2\sqrt{3}\gamma d^2\), containing an intermediate KT phase.

The present results are discussed in consistency with previous researches in literature:
The KT phase and hops of Josephson flux-line segments observed in the present simulations
provide a clear support to the scenario by Chakravarty et al. and Blatter et al.
\cite{Chakravarty,Blatter} formulated in order to explain the peculiar orientation independent
I-V characteristics and the power-law, non-Ohmic dissipations \cite{Iye,Ando}.
The recent experimental observation by Schilling et al. \cite{Schilling}
on first-order melting of Josephson flux-line lattice up to 10T for 
YBa\(_2\)Cu\(_3\)O\(_{7-\delta}\) is consistent with
\(B_{\rm mc}\simeq 50T\) for \(\gamma=8\).  The steep normal to
superconductivity phase boundary at high magnetic fields observed by Lundqvist et al.
\cite{Lundqvist} is able to be explained by the lower bound \(T^{\rm bare}_{\rm KT}\) on 
\(T_{\rm KT}(B)\).  Ooi and Hirata \cite{Ooi} detected a phase boundary
on which the 3D triangular lattice softens, very similar to the one in
our proposed phase diagram. 

Simulations are performed on the Numerical Materials Simulator
(SX-5) of NIMS. This study is partially supported by MEXT, Japan, under the Priority Grant 
No. 14038240.


\vskip1cm

\noindent Figure captions:

\noindent Fig. 1: Temperature dependence of in-plane helicity moduli for \(\gamma=8\)
and \(\gamma=20\).  The universal drop of helicity modulus for a pure 2D XY model at 
the bare KT transition is expected at \(\Upsilon\simeq 0.56J\) and 
\(T_{\rm KT}^{\rm bare}\simeq 0.89J/k_B\). 

\noindent Fig. 2: Structure factors at the transition point 
\(T_{\rm m, KT}=0.96J/k_B\) for \(\gamma=8\) (a) and \(\gamma=20\) (b).

\noindent Fig. 3: \(k_c\) profiles of the Bragg peaks at \([k_x,k_c]=[\pm 2f\pi/d,\pm \pi/d]\)
in Fig. 2.  The solid curve for \(\gamma=20\) is the result of the least-squares 
fitting to the Lorentzian function as described in text.

\noindent Fig. 4: \(k_x\) profiles of the Bragg spots at 
\([k_x,k_c]=[\pm 2f\pi/d,\pm \pi/d]\) for \(\gamma=20\) at several typical temperatures. 
The solid curves are results of the least-squares fittings to the power-law function
as described in text. 

\noindent Fig. 5: Temperature dependence of ratios of Josephson flux lines which 
contain segments hopping into neighboring block layers, of those which hop and/or
collide with neighbors in the same block layers, and populations of thermally excited, 
closed loops of Josephon vortices (normalized by \(40\times 240\)) and those
containing also pancake vortices (normalized by 240).

\noindent Fig. 6: \(B-T\) phase diagram for interlayer Josephson vortices with
a multicritical point.  The phase boundaries \(T_m(B), T_{\rm KT}(B)\) and \(T_{\times}(B)\) 
are associated with first-order, KT and 3D XY phase transitions as discussed in text. 

\end{document}